\documentclass[11pt]{article}
 
\usepackage{amsmath,amsthm}
\usepackage{amsthm}
\usepackage{amssymb}
\usepackage{amsfonts} 

\usepackage[latin1]{inputenc}

\usepackage{graphicx}

\newtheorem{theorem}{Theorem}

\newtheorem{proposition}{Proposition}

\def\be{\begin{equation}}
\def\ee{\end{equation}}
\def\bea{\begin{eqnarray}}
\def\eea{\end{eqnarray}}

\newcommand{\sect}[1]{\setcounter{equation}{0}\section{#1}}
\newcommand{\subsect}[1]{\subsection{#1}}

\newcommand{\bq}{\mathbf{q}}
\newcommand{\bp}{\mathbf{p}}

\newcommand{\hbq}{\hat{\mathbf{q}}}
\newcommand{\hbp}{\hat{\mathbf{p}}}
\newcommand{\hH}{\hat{\cal{H}}}
\newcommand{\hC}{\hat{C}}

\newcommand\Om\Omega

\newcommand{\bL}{\mathbf{L}}
\newcommand{\hbL}{\hat{\mathbf{L}}}

\newcommand{\cH}{{\cal{H}}}

\newcommand{\hq}{\hat{q}}
\newcommand{\hp}{\hat{p}}

\newcommand{\hr}{{r}}
\newcommand{\hte}{{\te}}

\newcommand{\la}{\eta}

\newcommand{\te}{\theta}
\newcommand{\dd}{{\rm d}}
\newcommand{\kk}{k}

 \def\RR{\mathbb{R}}

 \def\1{\'{\i}}
 \newcommand{\cM}{{\mathcal M}}
  \newcommand{\cR}{{\cal R}}

   \def\al{\alpha}

\def\rmi{{\rm i}}
\newcommand{\pd}{\partial}

\newcommand{\om}{\omega}
\newcommand{\De}{\Delta}

 \def\rme{{\rm e}}
\def\hamG{\hat{\cal H}_{\rm c}}

\def\NN{\mathbb N}

\def\ur{\varphi}
\def\ff{f}

\def\cY{\mathcal{Y}}
\def\SS{\mathbb S}
\newcommand{\LB}{{\rm c}}
\DeclareMathOperator\id{id}

\DeclareMathOperator\spec{spec}

\begin{document}

\
%\hfill\today
 \
 \smallskip
 
\vskip0.5cm

\noindent {\Large{\bf{An exactly solvable deformation of the     Coulomb  problem
\\[6pt] 
 associated with the Taub-NUT metric}}}

 \bigskip

\bigskip

\begin{center}
{\sc \'Angel Ballesteros$^1$,   Alberto Enciso$^2$,  Francisco J. Herranz$^1$,\\[4pt] Orlando Ragnisco$^3$ and Danilo Riglioni$^4$}
\end{center}

\noindent
{$^1$ Departamento de F\1sica,  Universidad de Burgos,
E-09001 Burgos, Spain\\ ~~E-mail: angelb@ubu.es,  fjherranz@ubu.es\\[10pt]
}
$^2$ Instituto de Ciencias Matem\'aticas,   CSIC, Nicol\'as Cabrera 13-15, E-28049 Madrid,
Spain\\ ~~E-mail: aenciso@icmat.es\\[10pt]
$^3$ Dipartimento di   Matematica e Fisica,  Universit\`a di Roma Tre and Istituto Nazionale di
Fisica Nucleare sezione di Roma Tre,  Via Vasca Navale 84,  I-00146 Roma, Italy  \\
~~E-mail: ragnisco@fis.uniroma3.it\\[10 pt]
$^4$ Centre de Recherches Math\'ematiques, Universit\'e de Montreal, H3T 1J4 2920 Chemin de la tour, Montreal, Canada\\
~~E-mail: riglioni@crm.umontreal.ca

\medskip

\begin{abstract}
\noindent 
In this paper we quantize the
 $N$-dimensional classical Hamiltonian system
$$
\cH= \frac{|\bq|}{2(\la + |\bq|)}\,\bp^2-\frac{\kk}{\la + |\bq|}  ,
$$
that can be regarded as a deformation of the Coulomb problem with coupling constant $k$, that it is smoothly recovered in the limit $\eta\to 0$. Moreover, the kinetic energy term in $\cH$ is just the one corresponding to an $N$-dimensional Taub-NUT space, a fact that makes this system relevant from a geometric viewpoint.
Since the Hamiltonian $\cH$ is known to be maximally superintegrable, we propose a quantization prescription that
preserves such superintegrability in the quantum mechanical setting. We
show that, to this end, one must choose as the kinetic part of the
Hamiltonian the conformal Laplacian of the underlying Riemannian
manifold, which combines  the usual Laplace--Beltrami operator on the
Taub--NUT manifold and a multiple of its scalar curvature. As a
consequence, we obtain a novel exactly solvable deformation of the
quantum Coulomb problem, whose spectrum is computed in closed
form  for positive values of $\eta$ and $k$, and showing that the well-known maximal degeneracy of the flat system is preserved in the deformed case. Several interesting algebraic and physical features of this new exactly solvable quantum system are analysed, and the quantization problem for negative values of $\eta$ and/or $k$ is also sketched.

\end{abstract}

\bigskip\bigskip\bigskip 



\newpage

\sect{Introduction}

Let us consider the two-parameter family of $N$-dimensional ($N$D) classical Hamiltonian systems given by
\be
\cH={\cal T}(\bq,\bp)+{\cal U}(\bq)=\frac{|\bq|}{2(\la + |\bq|)}\,\bp^2-\frac{\kk}{\la + |\bq|} ,
\label{aa}
\ee
where $\la$ and $\kk$ are real parameters, $\bq,\bp\in\RR^N$ are conjugate coordinates and momenta with canonical Poisson  bracket $\{q_i,p_j\}=\delta_{ij}$ and 
$$
\bq^2=\sum_{i=1}^N q_i^2,\qquad \bp^2=\sum_{i=1}^N p_i^2,\qquad |\bq|=\sqrt{\bq^2} .
$$
Clearly, the full Hamiltonian (\ref{aa}) can be regarded as an {\em $\la$-deformation} of the $N$D Euclidean  Coulomb  problem with coupling constant $\kk$, since the  limit $\la\to0$ yields
$$
{\cal H} =\frac12\, { \bp^2} -\frac{\kk}{ |\bq|} .
$$

Contemporarily, the system~\eqref{aa} can also be  interpreted as a Hamiltonian defined on a curved space, since the kinetic energy term ${\cal T}(\bq,\bp)$ provides the geodesic motion on the underlying $N$D  curved   manifold  $\cM=\RR^N\backslash\{\mathbf{0}\}$ with 
    metric 
\be
\dd s^2=\left(1+  \frac{\eta}{|\bq| }\right) \dd\bq^2,
\qquad |\bq|\neq 0,
  \label{metr}
\ee
and   scalar curvature   given by
\be
  R=\eta (N-1) \,\frac{  4(N-3) r+3\eta(N-2) }{ 4 r  (\eta+  r)^3},
  \label{ac}
\ee
where we have introduced the radial coordinate $r=|\bq|$. Note that the limit $\eta\to 0$ provides the flat/Euclidean expressions $\dd s^2=\dd\bq^2$ and $ R=0$. Therefore, we are dealing with a system defined on a conformally flat and spherically symmetric space $\cM$ with metric 
\be
\dd s^2= f(r)^2 \dd\bq^2, 
\label{aacc}
\ee
 whose conformal factor reads
\be
  f(r)= \sqrt{ 1+ \frac{\eta}{r }}\, .
  \label{ad}
   \ee

 It turns out that the mathematical and physical relevance of the Hamiltonian (\ref{aa}) relies on two important facts.    
On one hand, when $N=3$ the Hamiltonian  $\cH$ is directly related to a reduction~\cite{IK94} of the geodesic motion on the   Taub-NUT space~\cite{Ma82,AH85,GM86,FH87,GR88,IK95,uwano,BCJ,BCJM,GW07,JL}.
On the other hand,  it was shown in~\cite{sigma}  that (\ref{aa})
defines a maximally superintegrable  classical system, that is,   $\cH$ is
endowed with the maximum  possible number of $(2N-1)$ functionally
independent integrals of motion, all of which are, in this case,  {quadratic} in the momenta. 

In this paper we will present a quantization of~\eqref{aa} that preserves the maximal
superintegrability of the system. This result will provide an exactly
solvable deformation of  the Coulomb problem,
whose eigenvalue problem will be computed in detail with an emphasis on the cases
$\eta>0$ and  $k>0$. As expected, the spectrum of the standard $N$D Coulomb problem
will be recovered in the limit $\eta\to0$. 

It is worth stressing that the quantization of the Hamiltonian~\eqref{aa} is by no means straightforward, since the
kinetic energy term $\mathcal{T}({\mathbf{q}},{\mathbf{p}} )$
generates an ordering ambiguity when the classical position and
momenta are replaced by the corresponding operators. We shall see how
this problem can be solved by following the quantization procedure
proposed in~\cite{darbouxiii, annals} for another maximally
superintegrable quantum system on a different $N$D curved space: the
so-called {Darboux III oscillator system}, which is an exactly solvable
deformation of the harmonic oscillator potential that is associated
to the Darboux III space~\cite{Ko72, KKMW03}. Therefore, new exactly
solvable deformations of the  oscillator and   Coulomb problems
can be obtained when certain curved spaces with prescribed integrability properties are considered.
Moreover, we remark that the existence of additional integrals of the
motion associated with the maximal superintegrability of~\eqref{aa}, gives rise to an $\mathfrak{so}(N+1)$ Lie
symmetry algebra identical to the one underlying the $N$D Euclidean
Coulomb system. We will also show that this fact
makes it possible  to compute formally the discrete spectrum of the system in a very efficient way.

The structure of the paper is as follows. In the next section, the maximal
superintegrability of the classical Hamiltonian  $\cH$~\eqref{aa} is revisited (see~\cite{sigma, JPCS}). In
particular,  the geometric and dynamical  features of the
space~\eqref{metr} are studied as well  as  the connection between
$\cH$ and the  Taub-NUT metrics.  In Section 3 we present
a quantization for $\cH$ that preserves the full symmetry algebra of the
classical system and, therefore, its maximal superintegrability. Explicitly, we shall prove that this is achieved through the {\em conformal Laplacian quantization}~\cite{annals}, namely
\be
\hat{\mathcal{H}}_{\rm c}=-\frac{\hbar^2}2 \Delta_{\rm c}+\mathcal U= -\frac{\hbar^2}2 \left(\Delta_{\rm LB} -  \frac{  (N-2)}{4(N-1)} \,R
\right)+\mathcal U\, ,
\label{conf}
\ee
where $R$ is the scalar curvature, here given by (\ref{ac}), and 
 $ \Delta_{\rm c}$ is the conformal Laplacian~\cite{Baer}. Notice that
 the conformal Laplacian $\Delta_{\rm c}$ is the sum of the usual
 Laplace--Beltrami  operator $\Delta_{\rm LB}$  on the curved manifold
 ${\cal M}$ plus a multiple of the scalar curvature $R$ of  the
 manifold, while $\mathcal U$ is just the classical
 potential given in~\eqref{aa}  (see, e.g., the comprehensive
 reference~\cite{MRS} and~\cite{Wa84,  Landsman,  Liu}). 
 In order to prove this result we shall make use of the fact that~\eqref{conf}
can be related through a similarity transformation to the Hamiltonian obtained by means of the so-called {\em direct Schr\"odinger quantization} prescription~\cite{annals, JPCS}, namely
\begin{equation}
\label{direct}
\hat{\mathcal{H}} = -\frac{\hbar^2}{2 f(r)^2} \Delta  +\mathcal U\,   ,
\end{equation}
where $ f(r)= f(|\bq|)$ is the conformal factor of the metric (\ref{aacc}) and $\Delta$ is the Laplacian in the $\bq$ coordinates.
 The eigenvalue problem for these Hamiltonians will be rigorously
 solved in Section 4, where it is found that, for positive $k$ and $\eta$, the discrete spectrum
 of the system is a smooth deformation of the $N$D Euclidean Coulomb
 problem spectrum in terms of the parameter $\eta$ and, as expected,
 the quantum system presents the same maximal degeneracy as the $N$D hydrogen
 atom. The eigenvalue problem for other possible values of $k$ and $\eta$ are also sketched, and 
 the paper concludes with some remarks and open problems.

%%%%%%%%%%%%%%%%%%%%%%%%%%%%%%%%%%%%%%%%%%%%%%%%
 
\sect{The classical system}

The maximal superintegrability of the classical system $\cH$ is  explicitly stated through
 the following result~\cite{sigma}, that can be readily proven through direct computations.

  \begin{proposition}
  (i) The Hamiltonian $\cH$~\eqref{aa}       is endowed with      $(2N-3)$  angular momentum integrals  given by
($m=2,\dots,N$)
\be
  C^{(m)}=\!\! \sum_{1\leq i<j\leq m} \!\!\!\! ({q_i}{p_j} - {q_j}{p_i})^2 , \quad  
 C_{(m)}=\!\!\! \sum_{N-m<i<j\leq N}\!\!\!\!\!\!  ({q_i}{p_j} - {q_j}{p_i})^2 , \quad C^{(N)}=C_{(N)}  \equiv \bL^2,  \label{ba}
 \ee
where $ \bL^2$ is the square of the  total angular momentum.  

\noindent (ii)  The Hamiltonian $\cH$ Poisson-commutes with 
 the $\cR_{i}$ components ($i=1,\dots,N$) of the   Runge--Lenz $N$-vector given by
$$
\cR_{i}=\sum_{j=1}^N p_j ( q_j p_i - q_i p_j )+\frac{ q_i}{|\bq|} \left(\eta \cH+\kk\right).
$$
 \noindent
(iii) The set $\{ {\cal H},C^{(m)}, C_{(m)},  \cR_{{i}} \}$,      with $m=2,\dots,N$ and a fixed index $i$,    is formed by  $(2N-1)$ functionally independent functions. 
 
 \end{proposition}

Note that the following functional relation between the    Runge--Lenz vector ${\mathbf R}$, the angular momentum $\bL$ and the Hamiltonian $\mathcal{H}$ holds:
\be
{\mathbf R}^2=  \sum_{i = 1}^N \mathcal{R}_i^2 = 2 \bL^2 \mathcal{H}  + (\eta \mathcal{H}  + k)^2 .
\label{funct}
\ee
Therefore, Proposition 1 establishes that $\cH$ is a maximally superintegrable Hamiltonian that  is  endowed with  an  $\frak{so}(N)$  Lie--Poisson symmetry, since it is constructed  on a spherically symmetric space. Explicitly,  the functions $J_{ij}={q_i}{p_j} - {q_j}{p_i}$ with $i<j$ and $i,j=1,\dots,N$ span the $\frak
{so}(N)$ Lie--Poisson algebra   
$$
\{ J_{ij},J_{ik} \}= J_{jk} ,\qquad  \{ J_{ij},J_{jk} \}= -J_{ik} ,\qquad 
\{ J_{ik},J_{jk} \}= J_{ij} , \qquad i<j<k ,
$$
and   the   $(2N-3)$ angular momentum integrals $C^{(m)}$ and  $C_{(m)}$  (\ref{ba})  correspond to the quadratic Casimirs  of  some rotation subalgebras 
$\frak {so}(m)\subset \frak {so}(N)$. 

Moreover, if we also take into account  the additional integrals of
the motion $\cR_i$, then it can be   checked that $\cR_i$ and $J_{ij}$ span the Lie--Poisson algebra $\mathfrak{so}(N+1)$. In fact, we immediately obtain that
$$
\{ J_{ij}, \cR_k \}= \delta_{ik}\cR_j - \delta_{jk}\cR_i ,
$$
together with the quadratic Poisson bracket
$$
\{ \cR_i , \cR_j \} = -2 \mathcal{H}  J_{ij}.
$$
Nevertheless, if we now define 
\be
\tilde{J}_{0i} = \frac{\cR_i}{\sqrt{-2 \mathcal{H }}}, 
\qquad \tilde{J}_{ij} ={J}_{ij},
\label{clasA}
\ee 
we find that   the functions $\tilde{J}_{ij}$ close the Lie--Poisson algebra $\mathfrak{so}(N+1)$:
$$
\{ \tilde{J}_{ij},\tilde{J}_{ik} \}= \tilde{J}_{jk} ,\qquad  \{ \tilde{J}_{ij},\tilde{J}_{jk} \}= -\tilde{J}_{ik} ,\qquad 
\{ \tilde{J}_{ik}, \tilde{J}_{jk} \}= \tilde{J}_{ij} , \qquad i<j<k ,
$$
with $i,j,k=0,1,\dots,N$.
Therefore,  ${\cal H}$ turns out to be expressible as a function of the quadratic Casimir function for $\mathfrak{so}(N+1)$, since from~\eqref{funct} we have
\be
\sum_{i=1}^N \tilde{J}_{0i}^2 + {\mathbf{L}}^2 = - \frac{ (\eta \mathcal{H}  + k)^2}{2 \mathcal{H} }.
\nonumber
\ee

The Hamiltonian ${\cal H}$ can also  be expressed in terms of hyperspherical coordinates $r,\te_j$, 
  and canonical   momenta $p_r,p_{\te_j}$,   $(j=1,\dots,N-1)$  defined by
\be
q_j=r \cos\te_{j}     \prod_{k=1}^{j-1}\sin\te_k ,\quad 1\leq j<N,\qquad 
q_N =r \prod_{k=1}^{N-1}\sin\te_k ,
\label{bba}
\ee
so that
\be
r=|\bq|,\qquad \bp^2=p_r^2 + r^{-2} \bL^2,\qquad \bL^2=\sum_{j=1}^{N-1}p_{\te_j}^2\prod_{k=1}^{j-1}\frac{1}{\sin^{2}\te_k}.
\label{bbb}
\ee
Thus, for a given value of $\bL^2$, the Hamiltonian $\cH$ can be written as a 1D radial system:
\begin{equation}
\mathcal{H} (r, p_r) = \mathcal{T} (r, p_r) + \mathcal{U} (r) = \frac{r}{2(\eta + r)} \left( p_r^2 + \frac{{\mathbf L}^2}{r^2}\right) - \frac{k}{\eta + r} .
\label{xf}
\end{equation}

%%%%%%%%%%%%%%%%%%%%%%%%%%%%%%%%%%%%%%%%%%%%%%%%

\subsect{Geometric interpretation}

We stress that all the above Poisson algebraic results hold for {\em any} value of the deformation parameter $\eta$ and of the coupling constant $k$. Nevertheless,    $\cH$ comprises   {\em different}  classes of physical systems. In particular,    the domain   of the variable $r$ in $\cM$  depends on the sign of $\eta$:
\be
\eta>0:\quad r\in  (0,\infty) ;\qquad \eta<0:\quad  r\in ( |\eta|,\infty) .
\label{domain}
\ee
In particular, the case with $\eta>0$ turns out to be a system on a space:
\begin{itemize} 
\item With {\em positive} (nonconstant) scalar curvature if $N\geq 3$, which is given by~\eqref{ac}. Note that in this case the scalar curvature
  always diverges in the limit $r\to 0$, whereas it tends to zero for $r\to\infty$, as shown in Figure 1  for $N=3$.

\item With {\em negative} and finite (nonconstant) scalar curvature in the case $N=2$, where \eqref{ac} reads
$$
  R=-\frac{\eta}{(\eta+  r)^3},
$$
and whose $r\to 0$ limit is obviously $-{1}/{\eta^2}$ and whose
$r\to\infty$ limit also vanishes. This curvature is plotted in Figure 2.

\end{itemize}

It is worth noticing that there is a codimension-1 Riemannian
embedding of $\cM$ in Euclidean space. Specifically, let us denote by
$\dd \om^2$ the canonical metric on the sphere $\SS^{N-1}$. The metric
(\ref{metr}), which can be written as
\[
\dd s^2=\left(1+  \frac{\eta}{|\bq| }\right) \dd\bq^2= \left(1+  \frac{\eta}{r }\right) \left( \dd r^2 +r^2\dd\om^2 \right)
\]
is then recovered from the Euclidean metric 
\[
\dd x_1^2+\cdots+ \dd x_{N+1}^2
\]
in $\RR^{N+1}$ upon setting
\[
(x_1,\dots, x_N)=r\, \sqrt{1+\frac{\eta}{r}} \, \om\,,\qquad x_{N+1}=z(r)\,,
\]
where $\om=\om(\theta_1,\dots,\theta_{N-1})$ parametrizes a point in
$\SS^{N-1}$ via the hyperspherical coordinates used in  (\ref{bba}), $r$ takes values according to (\ref{domain})
  and the function $z(r)$ is defined as
\[
z(r)=\int_1^r \bigg(\frac{\eta(4 r' + 3\eta)}{4 r'
  (r'+\eta)}\bigg)^{\frac12}\, \dd r'\,.
\] 
For $N=2$, this embedding is represented in  Figure 3. The
fact that the surface is negatively curved and asymptotically flat is apparent
from these pictures.

%%%%%%%%%%%%%%%%%%%%%%%%%%%%%%%%%%%%%%%%%%%%%%%%

\begin{figure}
\begin{center}
\includegraphics[height=6.5cm]{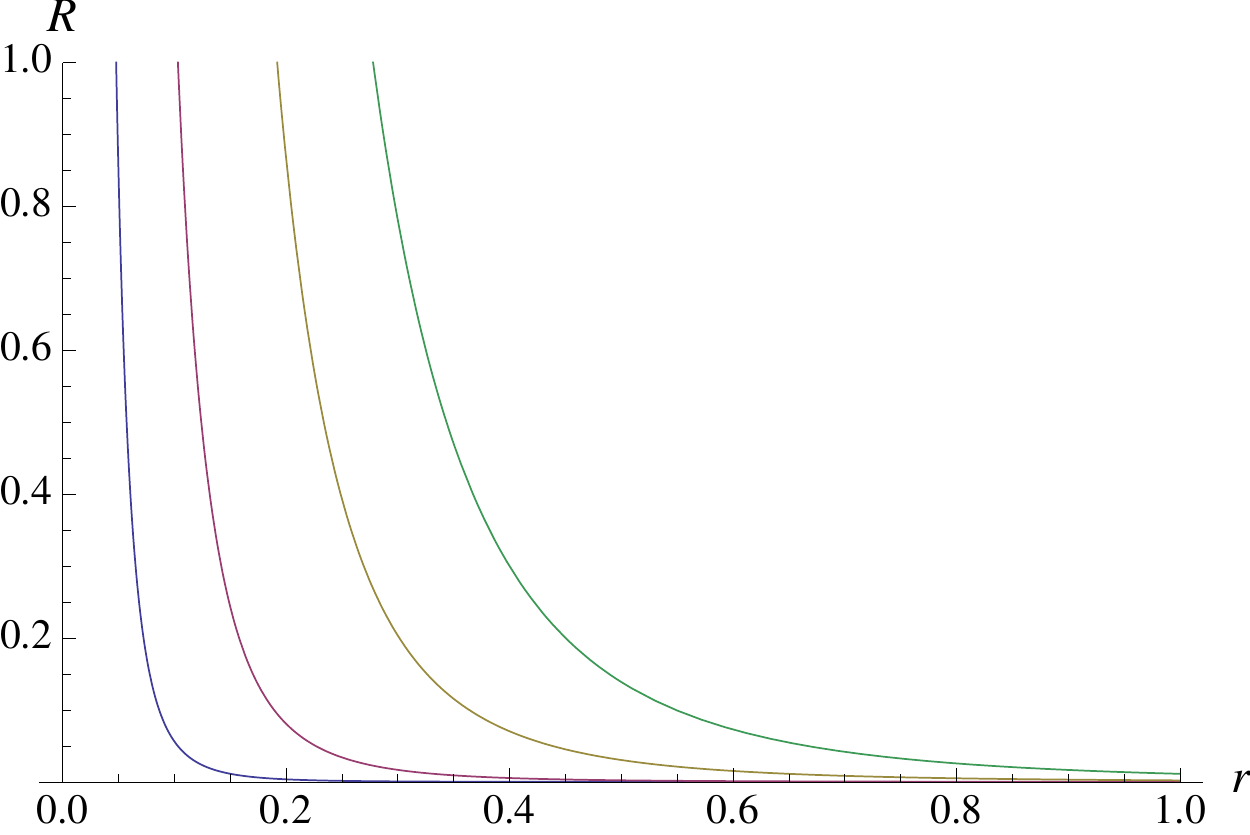}
\caption{Scalar curvature~\eqref{ac} of the Taub-NUT space  for $N=3$ and where $\eta=\{0.002, 0.01, 0.04,0.1\}$.
 \label{figure1}}
\end{center}
\end{figure}

%%%%%%%%%%%%%%%%%%%%%%%%%%%%%%%%%%%%%%%%%%%%%%%%

%%%%%%%%%%%%%%%%%%%%%%%%%%%%%%%%%%%%%%%%%%%%%%%%

\begin{figure}
\begin{center}
\includegraphics[height=6.6cm]{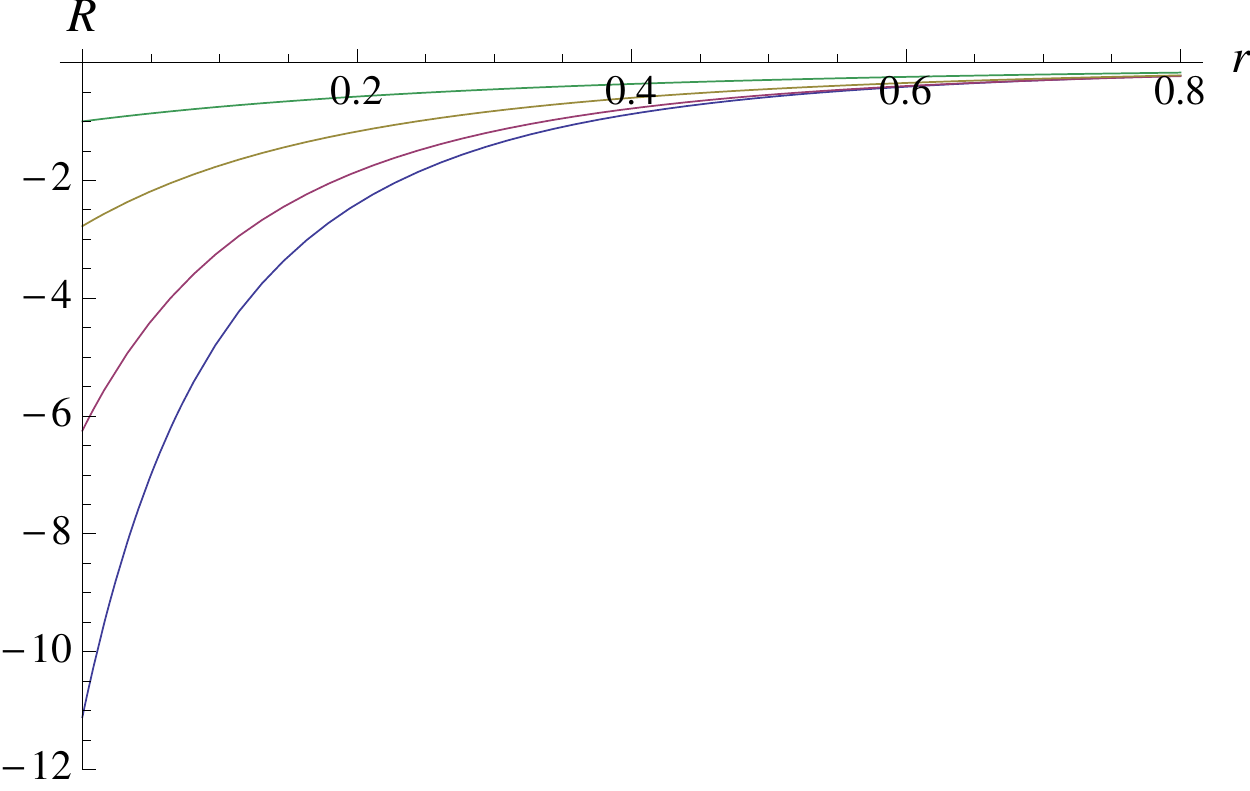}
\caption{Curvature~\eqref{ac}  of the Taub-NUT space   for $N=2$ and  with $\eta=\{0.3, 0.4, 0.6, 1\}$.
 \label{figure2}}
\end{center}
\end{figure}

%%%%%%%%%%%%%%%%%%%%%%%%%%%%%%%%%%%%%%%%%%%%%%%%

%%%%%%%%%%%%%%%%%%%%%%%%%%%%%%%%%%%%%%%%%%%%%%%%

\begin{figure}
\begin{center}
\includegraphics[height=11.2cm]{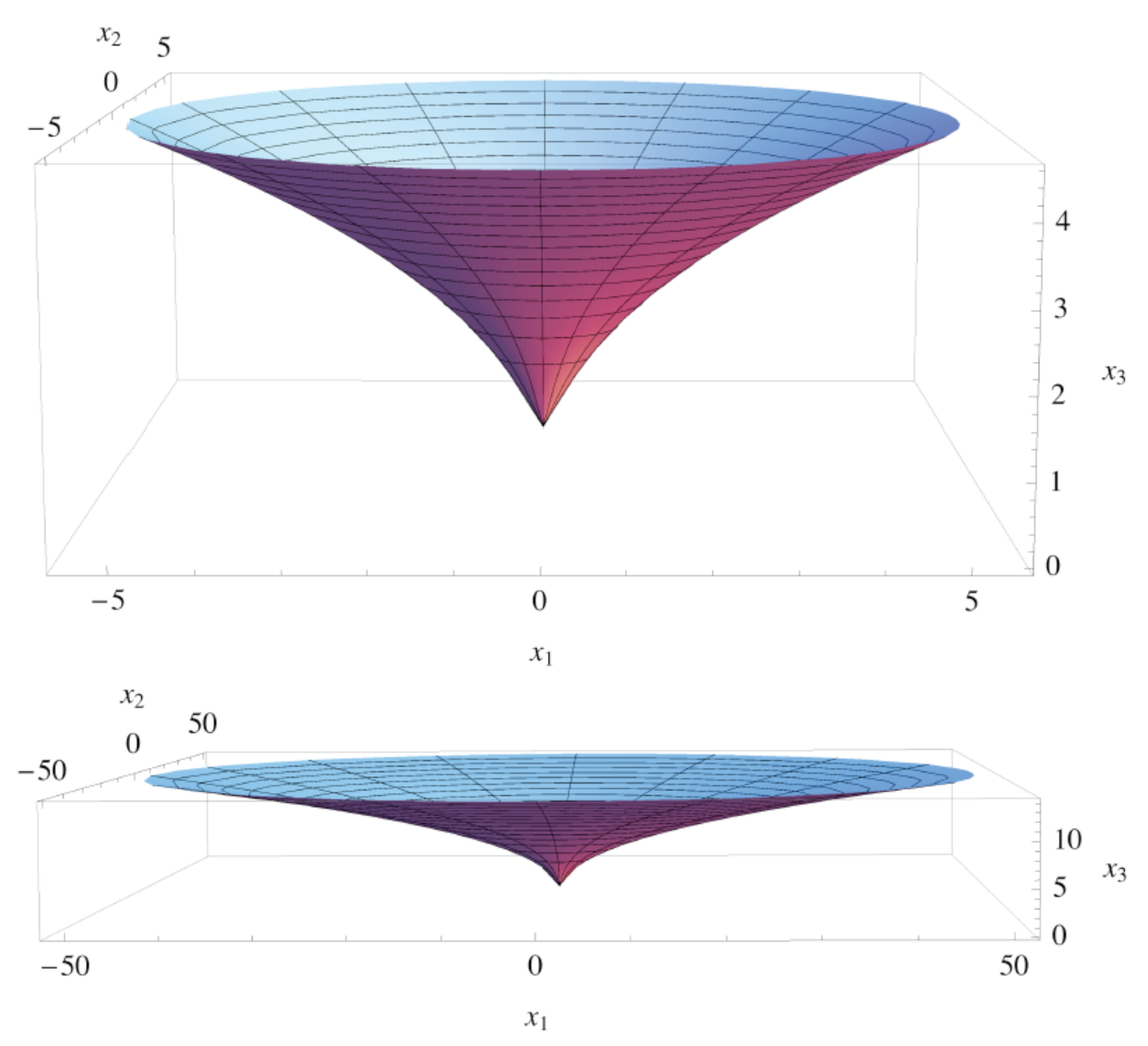}
\caption{3D Euclidean embedding of the Taub-NUT space for  $N=2$  with $\eta=1$ plotted for $r\in[0,5]$  and $r\in[0,50]$. 
 \label{figure3}}
\end{center}
\end{figure}

%%%%%%%%%%%%%%%%%%%%%%%%%%%%%%%%%%%%%%%%%%%%%%%%

 \newpage

\subsect{The connection with the Taub-NUT system}

It is important to stress that the case with $\eta>0$ and $k<0$ is the one related with  the Taub-NUT system studied in~\cite{IK94}, which describes the 3D reduction of the geodesic motion on the 4D Eucliden Taub-NUT metric.  In particular, let us write the Taub-NUT system $ {\cal H}_{\rm {T}}$  in the form~\cite{IK94,annals2}:
\bea
  && {\cal H}_{\rm {T}}=\frac {\bp^2}{2(1+4m/r)} +\frac{\mu^2}{2(4m)^2}\left( 1+\frac{4m}{r}\right)
   \nonumber\\
&&\qquad\qquad\quad =  
  \frac { r \bp^2 }{2( 4m +   r )}
+\frac{\mu^2 r/(4m)^2}{2( 4m + r  ) } 
+\frac{\mu^2 /(4m)}{4m +  r }  + \frac{\mu^2}{2 r ( 4m +   r )}.
\label{bd}
\eea
Next, if we consider the Hamiltonian $\mathcal{H}$~\eqref{aa} with 
\be
\eta =4m,\qquad \kk=-\frac{ \mu^2}{8m},
\nonumber
\ee
and we add a constant potential,   then
we get
$$
 {\cal H} +\frac{\mu^2}{2(4m)^2} =
  \frac { r \bp^2 }{2( 4m +   r )}
+\frac{\mu^2 r/(4m)^2}{2( 4m + r  ) } 
+\frac{\mu^2 /(4m)}{4m +  r }={\cal H}_{\rm {T}} -  \frac{\mu^2}{2 r ( 4m +   r )},
$$
thereby showing the equivalence between ${\cal H}$ and the Taub-NUT system (\ref{bd}) up to an additive term which is just a centrifugal
potential, that can be nevertheless reabsorbed by changing the value of the (conserved)
total angular momentum of the system (i.e., replacing $ \bL^2$ by
$\bL^2+\mu^2$ in (\ref{bbb})). 

It is worth recalling that the system $\mathcal{H}$~\eqref{aa} is
connected with the one introduced in~\cite{GM86} by considering the
asymptotic motion of monopoles when its separation is much greater
than their radii, which was already presented as a ``non-trivial deformation
of the Coulomb problem" in that reference. In what follows we will be mainly concerned with the case where $k$
and $\eta$ are both positive, whose spectral properties are
particularly interesting, but we will make some remarks about the
other choices of sign too.

%%%%%%%%%%%%%%%%%%%%%%%%%%%%%%%%%%

\subsect{Dynamical interpretation}

We shall   consider that $k>0$, in order to be able to
recover the hydrogen atom as the flat $\eta\to 0$ limit system, and $\eta>0$   ensuring that $r\in  (0,\infty) $. Note
that if $k<0$ the limiting case would be the repulsive Coulomb
problem, for which there are no bounded trajectories (or normalizable eigenfunctions).

\begin{figure}
\begin{center}
\includegraphics[height=6.3cm]{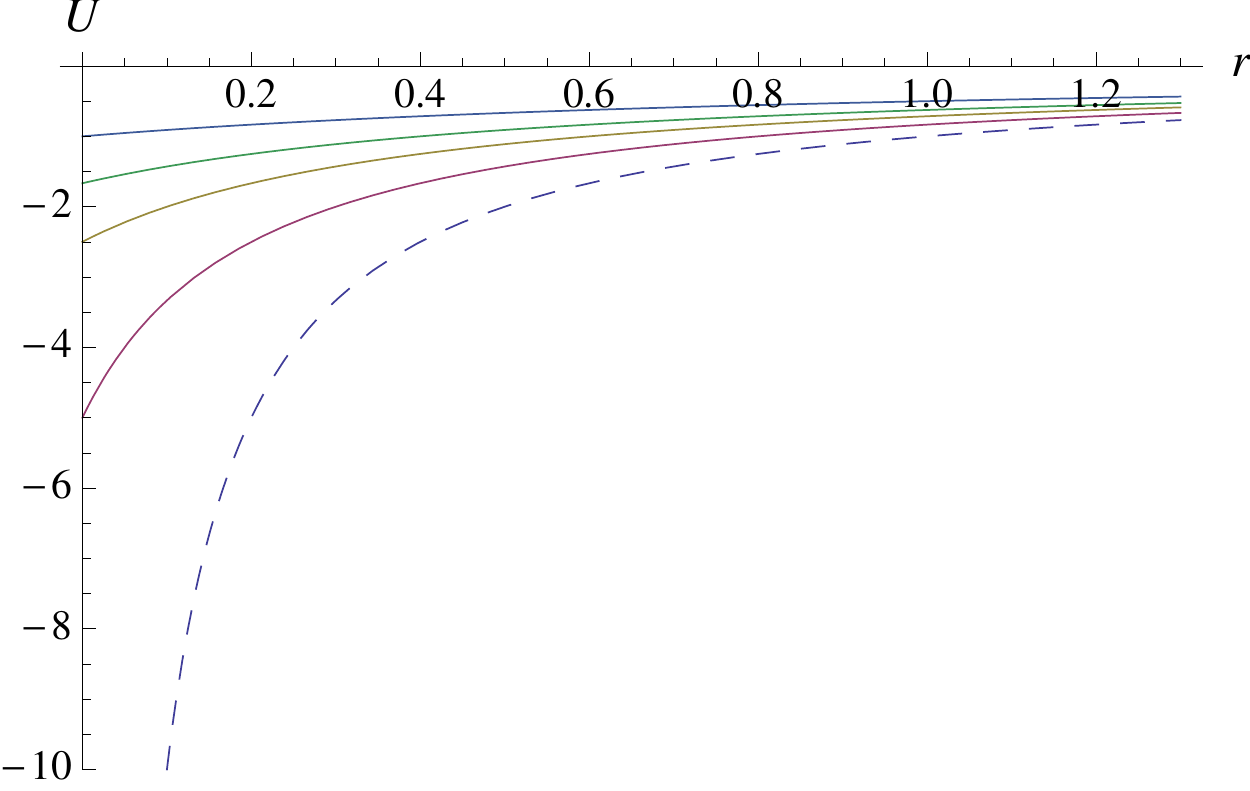}
\caption{The deformed Coulomb potential (\ref{pota}) for  $\eta=\{0,  0.2, 0.4, 0.6, 1\}$ with $k=1$. The  Euclidean  Coulomb potential ($\eta=0$) is represented by the dashed line.
 \label{figure4}}
\end{center}
\end{figure}

Unlike the standard Coulomb potential,  the ``deformed"     one  ${\cal U}(r)$ (\ref{xf})  is finite at $r=0$,  though both keep the same asymptotic  behavior  when $r\to \infty$, namely,
\be
{\cal U}(r)= - \frac{ \kk  }{\la+ r}    ,\qquad {\cal U}(0)=-\frac{\kk}\eta ,\qquad \lim_{r\to \infty}{\cal U}(r)=0 .
\label{pota}
\ee
This   potential is  shown in Figure  \ref{figure4}   for    several values of $\la$. In fact, the deformed Coulomb potential for a given $r$ is just the  flat  Coulomb one for a ``shifted" radial coordinate $r+\eta$.

\begin{figure}
\begin{center}
\includegraphics[height=6.3cm]{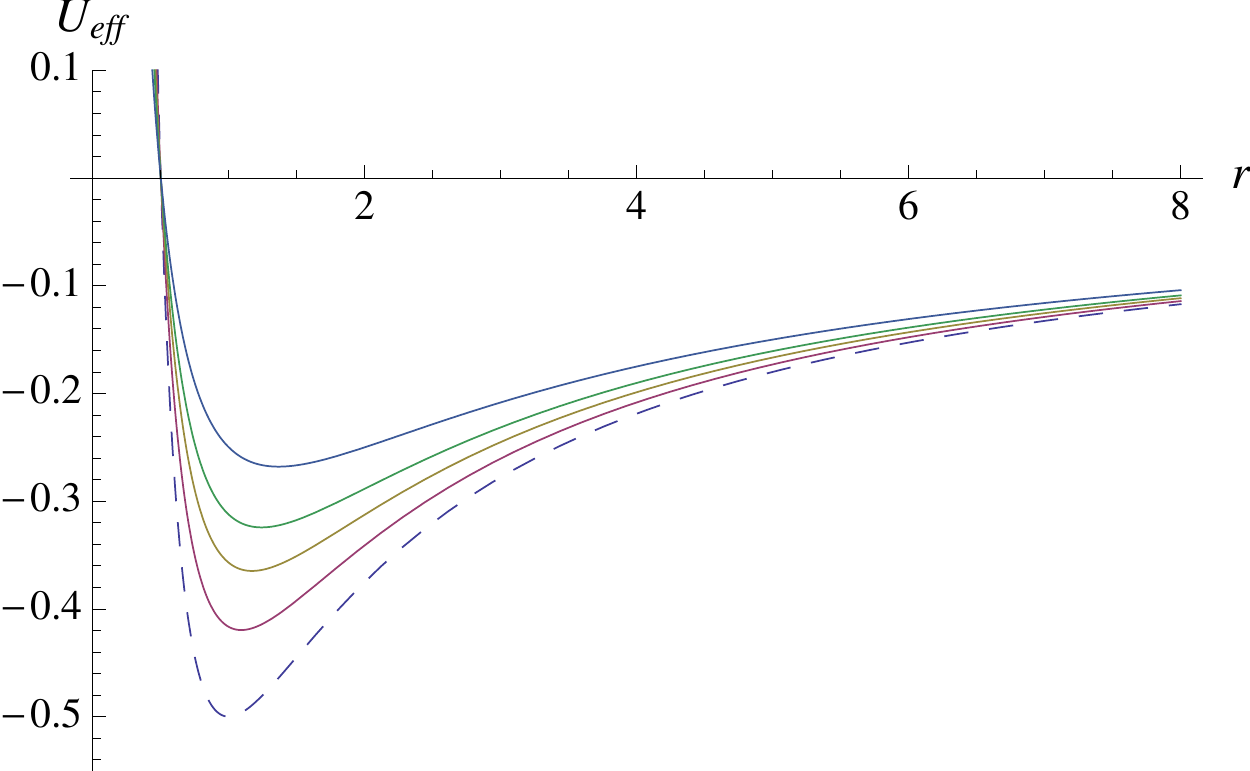}
\caption{The radial effective potential~\eqref{bbgg} for  $\eta=\{0, 0.2, 0.4, 0.6, 1\}$ with $\bL^2=1$ and $k=1$. The Euclidean  Coulomb effective potential ($\eta=0$) is represented by the dashed line.
 \label{figure5}}
\end{center}
\end{figure}

 However,
  since the underlying manifold  $\cM$  is non-flat,   the radial motion can be better understood by introducing a classical effective potential. In fact, let us consider the  canonical transformation defined by
\be
P(r,p_r)=\sqrt  {  \frac{r}{{\la+ r}}} \, p_r,  \qquad Q(r)=  \sqrt{ r(\la+ r)}+ \eta \ln  \left(\sqrt{r}+\sqrt{\eta+r } \right) ,
\label{bbhh}
\ee
such that  $ \{Q,P\}=1$ and $Q\in ( \eta\ln\sqrt{\eta} ,\infty)$. 
  In this way, we obtain that the radial Hamiltonian (\ref{xf})   is transformed into
  \be
\cH(Q,P)= \frac 12\, P^2+ {\cal U}_{\rm eff}(Q ),\qquad  {\cal U}_{\rm eff}(Q(r)  )= \frac{ \bL^2 }{2r(\la+ r)} -\ \frac{\kk} {\la+ r} .
 \label{bbgg}
 \ee
Consequently,   the radial motion for the classical system can be described as the one of a particle on $r\in  (0,\infty)$ under  the action of the effective potential $ {\cal U}_{\rm eff} $. This potential is represented in Figure~\ref{figure5}, where it can be appreciated that the effect of the $\eta$-deformation is to raise the minimum of  $ {\cal U}_{\rm eff}$ and to increase slightly the radius of the circular orbit for the system. Moreover, this classical effective potential does not depend on the dimension $N$ of the system. On the other hand, if the angular momentum vanishes the effective potential is just ${\cal U}(r)$ (\ref{pota}), depicted in Figure \ref{figure4}, which is never singular at the origin whenever $\eta\neq 0$. 

%%%%%%%%%%%%%%%%%%%%%%%%%%%%%%%%%%%%%%%%%%%%%%%%

%%%%%%%%%%%%%%%%%%%%%%%%%%%%%%%%%%%%%%%%%%%%%%%%

%%%%%%%%%%%%%%%%%%%%%%%%%%%%%%%%%%%%%%%%%%%%%%%%

%%%%%%%%%%%%%%%%%%%%%%%%%%%%%%%%%%%%%%%%%%%%%%%%

%%%%%%%%%%%%%%%%%%%%%%%%%%%%%%%%%%%%%%%%%%%%%%%%
 
\sect{Maximally superintegrable quantization}

Let us consider an $N$D    curved space  whose metric and associated classical kinetic term are given by
$$
{\rm d}s^2=\sum_{i,j=1}^N g_{ij}(\bq){\rm d}q_i{\rm d}q_j,\qquad {\cal T}(\bq,\bp)=\frac12\sum_{i,j=1}^N g^{ij}(\bq) {p_ip_j}\, .
$$
Then 
the corresponding Laplace--Beltrami   operator reads
\be
\Delta_{\rm LB}=\sum_{i,j=1}^N \frac 1{\sqrt{g}}\partial_i\sqrt{g} g^{ij}\partial_j ,
\label{LBop}
\ee
where $g^{ij}$ is the inverse of the metric tensor $g_{ij}$ and $g$ is the corresponding determinant. 
In principle, this operator could be used in order  to define the quantum kinetic energy operator in the form (see, for instance,~\cite{KKMW03,REFS1})
\be
\hat{\cal T}_{\rm LB}=-\frac {\hbar^2} 2 \Delta_{\rm LB} .
\nonumber
\ee
However, it turns out that a more popular quantization prescription used in the analysis of scalar field theories in General Relativity or when dealing with quantization on arbitrary Riemannian manifolds~\cite{MRS, Wa84, Landsman, Liu} is given by the so-called {\em conformal Laplacian} (or Yamabe operator)
\be
  \Delta_{\rm c} = \Delta_{\rm LB} -  \frac{  (N-2)}{4(N-1)} \,R     \, ,
  \nonumber
\ee
where $R$ is the scalar curvature of the underlying manifold. Note that both prescriptions coincide when either $N=2$ or $R=0$. Moreover, it was proven in~\cite{annals} for the Darboux III oscillator system, that this conformal Laplacian provides a maximally superintegrable quantization, since the quantum counterparts of all the constants of the motion of the classical system were explicitly found and, furthermore, the exact solvability of the system was obtained by making use of all these symmetries. 

In what follows, we present a maximally superintegrable quantization for the deformed Coulomb system $\cH$ (\ref{aa}), and we will explicitly obtain the full set of $(2N-2)$ algebraically independent  quantum observables that commute with the Hamiltonian. 
Hereafter, we will use the standard definitions for 
the quantum position $\hbq$ and momenta $\hbp$ operators:
$$
\hq_i \,\psi(\bq)=q_i\,\psi(\bq),\qquad \hp_i\,\psi(\bq)=-\rmi  \hbar \frac{\partial\,\psi(\bq)}{\partial q_i}, \qquad
[\hq_i,\hp_j]=\rmi \hbar \delta_{ij},
\qquad
i,j=1,\dots, N,
$$
together with the   conventions
$$
\nabla=\left(\frac\pd{\pd q_1},\dots,\frac\pd{\pd q_N}\right),\qquad \De=\nabla^2=\frac{\pd^2}{\pd^2 q_1}+\cdots +\frac{\pd^2}{\pd^2 q_N},\qquad \mathbf{q}\cdot\nabla =\sum_{i=1}^N  q_i\frac\pd{\pd q_i} .
$$
Note that the operator $|\hbq|$ is defined as $|\hbq|\,\psi(\bq)= |\bq|\,\psi(\bq)$.

 %%%%%%%%%%%%%%%%%%%%%%%%%%%%%%%%%%%%%%%%%%%%%%%%%%%
 
\subsect{Conformal Laplacian quantization}

It is straightforward to check that the Laplace--Beltrami operator (\ref{LBop})  for the manifold $\cM$  with metric tensor given by~(\ref{metr}) reads
\be
\Delta_{\rm LB}= \frac{ |\bq|}{\eta + |\bq|}\De - \frac{\eta(N-2)}{2|\bq| (|\bq|+\eta)^2}\, (\mathbf{q}\cdot\nabla ),
\nonumber
\ee
while the scalar curvature $R$ of the Taub-Nut space is given by~\eqref{ac}. With these two ingredients, a straightforward computation shows that the conformal Laplacian quantization of the classical deformed  Coulomb system $\cH$ (\ref{aa}) would be given by the Hamiltonian operator (\ref{conf}), namely
\be
{\hat{\cal H}}_{{\rm c} }= -\frac{\hbar^2}2\De_{\rm LB} +\hbar^2 \la (N-2)\,\frac{  4(N-3)|\bq|+3\la(N-2)  }{32|\bq|(\la+|\bq|)^3} -\frac{k }{\la + |\bq|} .
 \label{kn}
\ee

In order to prove that ${\hat{\cal H}}_{{\rm c} }$ is a maximally superintegrable quantization, the $(2N-2)$ algebraically independent operators commuting with $\hamG $ have to be explicitly found. By following~\cite{annals}, this can be achieved by considering the   Hamiltonian ${\hat{\cal H}}_{{\rm c} }$ in the form
 \be
\hat {\cal H}_{\rm c}=\rme^{\ff}\, \hat{\mathcal{H}}\,\rme^{-\ff},\qquad \ff(\bq) =   \frac{2-N}{4}  \,  \ln \left(1+ \frac{\la}{|\bq|} \right) ,
 \label{transf}
  \ee
which is provided by a similarity transformation from the so-called ``direct Schr\"odinger  quantization" prescription (\ref{direct})  for the deformed Coulomb system:
 \be
\hat{\mathcal{H}}  =  \frac{     -\hbar^2|{\mathbf q}|}   {2 (\eta + {|\mathbf q}|)} \,\De - \frac{  k}{ \eta+{|\mathbf q}|}  .
 \label{ca} 
 \ee
With this aim in mind,  we firstly  recall (see~\cite{JPCS})   the   superintegrability  properties of the Hamiltonian $\hH$  (\ref{ca})  which are  summarized in the following statement, and that are worth to be compared with Proposition~1.

  \begin{proposition}
    Let   $\hH$ be the quantum Hamiltonian given by~\eqref{ca}.
Then:

\noindent
(i) $\hH$ commutes with the   $(2N-3)$ quantum angular momentum operators ($m=2,\dots,N$):
\be
  \hC^{(m)}=\!\! \sum_{1\leq i<j\leq m} \!\!\!\! ( \hq_i \hp_j- \hq_j \hp_i)^2  , \quad \ 
  \hC_{(m)}=\!\!\! \sum_{N-m<i<j\leq N}\!\!\!\!\!\!  ( \hq_i \hp_j- \hq_j \hp_i)^2  ,  \quad\  \hC^{(N)}=\hC_{(N)}=\hat \bL^2 ,
  \label{cb}
 \ee
 where $\hat \bL^2$   is the total quantum angular momentum operator, as well as with the $N$ Runge--Lenz operators given by $(i=1,\dots,N)$:
 \be
  \hat{\cal R}_{i}  = \frac{1}{2}\sum_{j=1}^N \hp_j ( \hq_j \hp_i - \hq_i \hp_j )+ \frac{1}{2} \sum_{j=1}^N  ( \hq_j \hp_i - \hq_i \hp_j )\hp_j+\frac{ \hq_i}{|\hbq|} \left(\eta \hH+\kk\right).
\label{cc}
\ee

 \noindent
(ii) Each of the three  sets $\{{\hH},\hC^{(m)}\}$,  
$\{{\hH},\hC_{(m)}\}$ ($m=2,\dots,N$) and   $\{    \hat{\cal R}_{i} \}$ ($i=1,\dots,N$) is  formed by $N$ algebraically independent  commuting operators.

 \noindent
(iii) The set $\{ { \hH},\hC^{(m)}, \hC_{(m)},    \hat{\cal R}_{i} \}$ for $m=2,\dots,N$ with a fixed index $i$    is  formed by $(2N-1)$ algebraically independent operators. 

 \noindent
(iv) $\hH$ is formally self-adjoint on the Hilbert space  of square-integrable functions with respect to the the scalar product
\be
\langle \Psi | \Phi \rangle = \int_{\RR^N} \overline{{\Psi}(\bq)} \Phi(\bq)\left(1+\frac{\la}{|\bq|}\right){\dd}\bq .
\label{product1}
\ee
\end{proposition}

With this result at hand, the maximal superintegrability of the conformal Laplacian quantization $\hamG$ can be   obtained by making use of the similarity transformation~\eqref{transf} in order to get the quantum integrals for $\hamG$ starting from the ones for $\hH$. More explicitly, we have:

 \begin{proposition} (i) The quantum Hamiltonian   $\hamG$ given by  \eqref{kn}    commutes with the  operators \eqref{cb}   as well as with 
  the following $N$ operators of  Runge--Lenz type  $(i=1,\dots,N)$:
  \bea
&&  \hat{\cal R}_{{\rm c},i}= \frac{1}{2}\sum_{j=1}^N\left(  \hp_j + \rmi  \hbar \, \eta\, \frac{(N-2)\hq_j}{4
(\eta+|\hbq|)\, \hbq^2}  \right) ( \hq_j \hp_i - \hq_i \hp_j )\nonumber\\[2pt]
&&\qquad\qquad  + \frac{1}{2} \sum_{j=1}^N  ( \hq_j \hp_i - \hq_i \hp_j ) \left(  \hp_j + \rmi  \hbar \, \eta\, \frac{(N-2)\hq_j}{4
(\eta+|\hbq|)\,\hbq^2}  \right)   +\frac{ \hq_i}{|\hbq|} \left(\eta \hH_{\rm c}+\kk\right).
\nonumber
\eea
 
 \noindent
(ii) Each of the three  sets $\{{\hamG},\hC^{(m)}\}$,  
$\{{\hamG},\hC_{(m)}\}$ ($m=2,\dots,N$) and   $\{  \hat{\cal R}_{{\rm c},i}\}$ ($i=1,\dots,N$) is  formed by $N$ algebraically independent  commuting operators.

\noindent
(iii) The set $\{ { \hamG},\hC^{(m)}, \hC_{(m)},   \hat{\cal R}_{{\rm c},i}\}$ for $m=2,\dots,N$ with a fixed index $i$    is  formed by $(2N-1)$ algebraically independent operators.

\noindent
(iv) $\hamG$ is formally self-adjoint on the Hilbert space $L^2(\cM)$
with its natural scalar product 
\be
\langle \Psi | \Phi \rangle_{\rm c} = \int_{\RR^N} \overline{{\Psi}(\bq)}\, \Phi(\bq)\, 
\left(1+\frac \eta{|\bq|} \right)^{N/2}\,\dd\bq .
\nonumber
\ee
\end{proposition}

Notice that the multiplication operator $\rme^f$ (\ref{transf})  also
defines a unitary transformation mapping 
$$
L^2\left(\RR^N,(1+ {\la}/{|\bq|})\,\dd \bq\right)\quad {\rm into}\quad L^2\left(\RR^N,(1+ {\la}/{|\bq|})^{N/2}\dd\bq\right),
$$
 which is the natural $L^2$ space defined by the Riemannian metric. Therefore, according to the above statement, $\hamG$ can be seen as an  appropriate quantization of the classical Hamiltonian (\ref{aa}), as it manifestly preserves the maximal superintegrability of the classical   system.

 %%%%%%%%%%%%%%%%%%%%%%%%%%%%%%%%%%%%%%%%%%%

\subsect{Radial Schr\"odinger equation}

By mimicking the same approach presented in~\cite{annals}, it is straightforward to prove that  the quantum radial Hamiltonian corresponding to (\ref{ca}) is 
  \be
 {\hH}=  \frac{\hr}{2(\la + \hr)}\left( \frac{1}{\hr^{N-1}}\, \hp_r\, \hr^{N-1} \,\hp_r+\frac{\hbL^2}{\hr^{2}}-\frac{2 k}{\hr} \right)    ,
 \label{lg}
 \ee
 where $\hbL^2$ is the square of the total  quantum  angular momentum  operator,  given by
\be
\hbL^2=\sum_{j=1}^{N-1}\left( \prod_{k=1}^{j-1}\frac{1}{\sin^{2}\hte_k}  \right) \frac{1}{(\sin\hte_j)^{N-1-j}}\, \hp_{\te_j}(\sin\hte_j)^{N-1-j} \,\hp_{\te_j}.
\nonumber
\ee
Here the   hyperspherical coordinates  (\ref{bba})  have been used    together with 
\be
  \hp_r= -\rmi\hbar\frac{\partial}{\partial r} ,
  \qquad \hp_{\te_j}=-\rmi\hbar \frac{\partial}{\partial{\te_j}},\qquad j=1,\dots, N-1.
  \label{diffop}
\ee

After reordering terms and  by introducing  the differential operators~\eqref{diffop}  within the Hamiltonian (\ref{lg}), we arrive at the following Schr\"odinger equation,  
      \be
\frac{r}{2( \la + r)}\left(  {-\hbar^2}\partial_r^2 -\frac{\hbar^2(N-1)}{r}\,\partial_r + \frac{\hat{\mathbf{L}}^2}{r^2}-\frac{2k}{r}  \right)\Psi(r, \boldsymbol{\te}) = E\,\Psi(r,{\boldsymbol{\te}}) ,
 \label{lh}
 \ee
where $\boldsymbol{\te}=(\te_1,\dots,\te_{N-1})$.  By taking into account that the hyperspherical harmonics $Y(\boldsymbol{\te})$ are such that
\be
\hbL^2 Y(\boldsymbol{\te})= \hat{C}_{(N)} Y(\boldsymbol{\te})=   \hbar^2 l(l+N-2)\,Y(\boldsymbol{\te}),\quad l=0,1,2\dots  ,
\nonumber
\ee
where $l$ is the angular momentum quantum number, the equation~\eqref{lh}
admits a complete set of factorized solutions of the form
\be
\Psi(r, \boldsymbol{\te})= \Phi(r)Y(\boldsymbol{\te}),
\nonumber
\ee
and, moreover, 
\be
\hat{C}_{(m)}\Psi = c_{m} \Psi,\quad m=2,\dots , N,
\nonumber
\ee
where the eigenvalues  $c_m$ of the operators  $\hat{C}_{(m)}$ (\ref{cb}) are related to the  $(N-1)$ quantum numbers of the angular observables in the form 
\be
c_{k}\leftrightarrow l_{k-1},\quad k=2,\dots, N-1,\qquad c_N \leftrightarrow l .
\nonumber
\ee
Therefore, we can write
\be
Y(\boldsymbol{\te})\equiv Y^{c_N}_{c_{N-1},..,c_2}(\theta_1,\theta_2,...,\theta_{N-1}) \equiv Y^{l}_{l_{N-2},..,l_1}(\theta_1,\theta_2,...,\theta_{N-1}) ,
\nonumber
\ee
and the radial Schr\"odinger equation provided by $\hat{\cal H}$  reads
\be
\frac{r}{2(\la + r)} \left( {-\hbar^2}\left(\frac{\dd^2}{\dd r^2} +\frac{N-1}{r}\frac{\dd}{\dd r} -\frac{l(l+N-2)}{r^2}\right) -\frac{2k}{r} \right) \Phi(r) = E\, \Phi(r) .
\label{sa}
\ee

Since the radial Hamiltonians ${\hat {\cal H}}$ and $\hamG$   are   related through the unitary transformation (\ref{transf}), namely,
$$
\hamG=\left(1+\frac{\la}{r} \right)^{(2-N)/4}\,  {\hat {\cal H}} \, \left(1+\frac{\la}{r} \right)^{(N-2)/4},
$$
 the radial equation coming from the conformal Laplacian quantization (\ref{kn})  is found to be
\bea
&&\left\{  - \frac{\hbar^2 r}{2(  \la+r) } \left( \frac{\dd^2}{\dd r^2} +\left( \frac{N-1}{r}   -\frac{\la(N-2) }{2r( \la+r) } \right) \frac{\dd}{\dd r} -\frac{l(l+N-2)}{r^2}\right) \right.\nonumber\\
 &&\qquad\quad \left. -\frac{k}{  \la+r } +\hbar^2\la (N-2)     \frac{ 4 (N-3) r+ 3\la(N-2) }{32r(\la  +r)^3} \right\}\Phi_{\rm c}(r)= E\, \Phi_{\rm c}(r)\,.
 \label{ssaa}
 \eea
Therefore, the two radial equations (\ref{sa}) and (\ref{ssaa})  will share the same energy spectrum and their radial wave functions will be related through
\be
\Phi_{\rm c}(r)=\left(1+\frac{\la}{r} \right)^{(2-N)/4}\, \Phi(r).
 \label{lk}
\ee

 %%%%%%%%%%%%%%%%%%%%%%%%%%%%%%%%%%%%%%%%%%%%%%%%%%%
 
 \sect{Spectrum and eigenfunctions}

In this section we shall compute, in a rigorous manner,   the
(continuous and discrete) spectrum and eigenfunctions of the quantum
Hamiltonian $\hamG$ (\ref{kn}). Although we will focus on the case $\la>0$ and
$k>0$, which is particularly interesting due to its connection with the Coulomb problem, we will also comment on the
remaining possibilities for the signs of $\la$ and $k$.

To begin with, let us observe that the spherical symmetry of $\hamG$ (\ref{kn})  leads us to decompose
\begin{equation}\nonumber
L^2(\cM)=\bigoplus_{l\in\NN}L^2(\RR^+,\dd\nu)\otimes\cY_l\,,
\end{equation}
where $\dd\nu=r^{N-1}(1+\la /r) 
^{N/2}\dd r$ and $\cY_l$ is the finite-dimensional space of (generalized) spherical harmonics, defined by
\[
\cY_l:=\big\{Y\in L^2(\SS^{N-1}):\De_{\SS^{N-1}}Y=-l(l+N-2)Y\big\}\, ,
\]
where $\NN$ stands for the set of nonnegative integers and
$\De_{\SS^{N-1}}$ denotes the Laplacian on the $(N-1)$D sphere
$\SS^{N-1}$. We recall that $\cY_l$ consists of the restriction to the
sphere of the harmonic polynomials in $\RR^N$ that are homogeneous of
degree $l$.

This decomposition implies that the wave function can be written as
\[
\Psi_\LB(\bq)=\sum_{l\in\NN}Y_l(\boldsymbol{\theta})\,\Phi_{\LB,l}(r)\,,
\]
with $\boldsymbol{\theta}=\bq/r\in\SS^{N-1}$, $r=|\bq|$ and
$Y_l\in\cY_l$.

In view of the expression for the radial effective potential (see (\ref{se}) below), it is
not hard to see that $\hamG$ can be regarded as a densely defined
self-adjoint operator on $L^2(\cM)$ that, by virtue of the above
decomposition, can be written as  
\begin{equation}\nonumber
\hamG=\bigoplus_{l\in\NN} \hat H_{\LB,l}\otimes \id_{\cY_l}\,.
\end{equation}
From the expression of the potentials we see that each
operator $\hat
H_{\LB,l}$, 
\begin{multline}
  \hat H_{\LB,l}=  - \frac{\hbar^2 r}{2(  \la+r) } \left( \frac{\dd^2}{\dd r^2} +\left( \frac{N-1}{r}   -\frac{\la(N-2) }{2r( \la+r) } \right) \frac{\dd}{\dd r} -\frac{l(l+N-2)}{r^2}\right) \\
 \quad  -\frac{k}{  \la+r } +\hbar^2\la (N-2)     \frac{ 4 (N-3) r+ 3\la(N-2) }{32r(\la  +r)^3} ,
 \label{sscc}
\end{multline}
can be taken to be the Friedrichs extension of the above differential
operator acting on the space of smooth, compactly supported functions
$C^\infty_0(\RR^+)$ (see e.g.~(\ref{ssaa})). Therefore, the spectrum
of $\hamG$ is just
\[
\spec(\hamG)=\overline{\bigcup_{l\in\NN}\spec(\hat H_{\LB ,l})}\,.
\]

In order to analyze the spectrum, the quantum effective potential $\hat {\cal U}_{{\rm eff},l}$ will be helpful. This can be obtained by applying 
 the change of radial variable $Q=Q(r)$ given by (\ref{bbhh}) together with  a transformation of the radial wave function  $\Phi_{\LB,l}(r) \mapsto \ur(Q(r)) $ and by imposing that these transformations  map the Schr\"odinger equation $\hat H_{\LB ,l}\Phi_{\LB,l}=E\, \Phi_{\LB,l}$ into  
\bea
 \left( - \frac {\hbar^2}2 \frac{\dd^2}{\dd Q^2} + \hat {\cal U}_{{\rm eff},l}(  Q )\right) \ur(Q)=E\, \ur(Q) .
\nonumber
\eea
This requires to introduce
\bea
\Phi_{\LB,l}(r) =\frac{r^{(1-N)/2}}{(1+\frac{\la}{r})^{(N-1)/4}}\, \ur(Q)
\nonumber
\eea
in the  radial Schr\"odinger equation provided by the Hamiltonian (\ref{sscc}),  thus yielding 
 \be
 \hat {\cal U}_{{\rm eff},l}(r  )= \frac{r}{2(\eta+r)} \left( -\frac{\hbar^2 (\eta^2+4r^2)}{16r^2(\eta+r)^2} + \frac{\hbar^2}{r^2}\left( l(l+N-2) + \frac{(N-2)^2}{4}\right) - \frac{2k}{r}  \right).
 \label{se}
\ee

The behavior of the resulting quantum effective potential   deserves
some comments since, in contrast to the classical case,  $\hat {\cal U}_{{\rm eff},l}(r)$ depends on the dimension $N$ of the Taub-NUT manifold:

\begin{itemize}

\item For $N\geq 3$, the behavior of the effective potential is always similar to that of the classical one~\eqref{bbgg} (see figure~\ref{figure5}), irrespectively of the value of $l$. In particular, even in the case that $l=0$,  the quantum effective potential  is such that
 \be
\lim_{r\to 0} \hat{\cal U}_{{\rm eff},l}(r)=+\infty,\qquad \lim_{r\to \infty} \hat{\cal U}_{{\rm eff},l}(r)=0   .
\label{bzz}
\ee
Moreover, $ \hat {\cal U}_{{\rm eff},l}$ has always a unique minimum at $r_{\rm min}$ 
whose $\la\to 0$ non-deformed Coulomb limit is given  by
\bea
  \la=0:&&  r_{0,\rm min} = \frac{ \hbar^2}{k}  (l(l+N-2) +(N-1)(N-3)/4),\nonumber\\
  &&    \hat{\cal U}_{{\rm eff},l}(r_{0,\min} )=-\frac{2 k^2}{\hbar^2(4l(l+N-2) +(N-1)(N-3))} ,
\nonumber
 \eea
 with the exception of the non-deformed case with $N=3$ and $l=0$ which gives $ r_{0,\rm min} = 0$ and $ \hat{\cal U}_{{\rm eff},l}(r_{0,\min} )\to -\infty$.

\item When $N=2$, the effective potential (\ref{se}) reduces to
 \be
 \hat {\cal U}_{{\rm eff},l}(r  )= \frac{r}{2(\eta+r)} \left( -\frac{\hbar^2 (\eta^2+4r^2)}{16r^2(\eta+r)^2} + \frac{\hbar^2\,l^2}{r^2} - \frac{2 k}{r}  \right).
\nonumber
\ee
If $l=0$ we obtain that
\be
\lim_{r\to 0} \hat {\cal U}_{{\rm eff},0}(r)= -\infty ,
\qquad \ 
\lim_{r\to \infty} \hat {\cal U}_{{\rm eff},0}(r)= 0 ,
\nonumber
\ee 
so that $ \hat {\cal U}_{{\rm eff},0}$ has no local minima and  always takes  negative  values. On the contrary, for $l\geq 1$ we find   
that $ \hat {\cal U}_{{\rm eff},l}$ has  the
  same limiting values (\ref{bzz})  and  has  only one local minimum,
  where the effective potential is negative.

\end{itemize}

 %%%%%%%%%%%%%%%%%%%%%%%%%%%%%%%%%%%%%%%%%%%%%%%%%%%
 
 \subsect{The case $k>0$, $\eta>0$}

Let us now compute the discrete eigenvalues and eigenfunctions of
$\hamG$ when $k$ and $\eta$ are positive. To begin with, let us recall
that, with $k>0$, the radial part $\psi_{n,l}(r)$ of the
eigenfunctions for the standard $N$D Coulomb problem satisfies the following
Schr\"odinger equation corresponding to a spherical harmonic of degree
$l$:
\be
 \left(-\hbar^2 \left( \frac{\dd^2}{\dd r^2} + \frac{N-1}{r}\frac{\dd}{\dd r} \right)+\frac{\hbar^2 l(l+N-2)}{r^2} - \frac{2 k}{r} \right)\psi_{n,l}(r)=  2E_{n,l}^0\, \psi_{n,l}(r) \,,
 \label{coul}
\ee 
with eigenvalues
\be
E_{n,l}^0=-\frac{k^2}{ 2 \hbar^2 \left (n+l+\frac{N-1}{2} \right)^2}\,.
\label{energy}
\ee
The explicit expression of $\psi_{n,l}$ is given, in terms of generalized Laguerre polynomials,  by
\begin{equation}\label{Psi}
\psi_{n,l}(r)=r^l \exp\left( - \frac{kr}{ \hbar^2 \left(n+l+\frac{N-1}{2} \right)}  \right)   \,L_n^{2l+N-2}\left( \frac{2kr}{\hbar^2\left (n+l+\frac{N-1}{2} \right)}\right) ,
\end{equation}
up to the corresponding normalization constant.

Now, the solution for the eigenvalue problem of the deformed Coulomb Hamiltonian $\hat{\mathcal{H}}$  can be obtained if we realize that~\eqref{sa} can be rewritten in the form
\be
\left( {-\hbar^2}\left(\frac{\dd^2}{\dd r^2} +\frac{N-1}{r}\frac{\dd}{\dd r} -\frac{l(l+N-2)}{r^2}\right) -\frac{2k}{r} \right) \Phi(r) = 2E\, \left(
1 +\frac{\la}{r} \right)\,\Phi(r) ,
\label{sab}
\ee
which is nothing but the equation~\eqref{coul}
\be
\left(-\hbar^2 \left( \frac{\dd^2}{\dd r^2} + \frac{N-1}{r} \frac{\dd}{\dd r} \right)+\frac{\hbar^2 l(l+N-2)}{r^2} - \frac{2K}{r} \right)\Phi(r)= 2E\, \Phi(r) ,
\nonumber
\ee
provided that  we set
\be
K = k + \la\, E .
\label{formK}
\ee
This means that the eigenvalue problem for the  Hamiltonian
$\hat{\mathcal{H}}$ is formally the standard Coulomb problem with a new
coupling constant $K$ that depends on the initial coupling constant
$k$, the energy $E$ and the $\la$ parameter. This fact does not immediately
yield the eigenvalues of the problem, however, because the
integrability (and boundary) conditions that one must impose are
different in this case. Note that this is analogous to what happens with the quantum harmonic oscillator eigenvalue problem
with respect to the superintegrable oscillator defined on the Darboux III curved space~\cite{annals}. 
Consequently, we shall directly solve the equation~\eqref{sab} and next analyze the corresponding results.

 For $E<0$, it can be shown that  the general solution of \eqref{sab} must be of
the form
\be\label{Phic1c2}
\Phi(\rho)=\rho^l \rme^{-\rho/2}\big(c_1U_{-\al}^{\beta}(\rho)+c_2 M_{-\al}^{\beta}(\rho)\big)\,,\
\ee
where $U^{\beta}_{-\alpha}(\rho)$  and $M^\beta_{-\alpha}(\rho)$  stand for two independent confluent hypergeometric functions (called, respectively,  Tricomi and
Kummer functions)  and we have set
\be\label{defs}
\rho=\frac{2r\sqrt{-2E}}\hbar\,,\qquad \al=\frac {k+\eta E}{\hbar\sqrt{-2E}} -l-\frac{N-1}2 , \qquad \beta =2l+N-1 \, .
\ee

Let us recall that the large--$\rho$ asymptotic behavior of the above functions is
\[
U^{\beta}_{-\alpha}(\rho)\sim \rho^{\alpha}[1+o(1)]\,,\qquad M^\beta_{-\alpha}(\rho)\sim\rme^\rho
\rho^{-\alpha-\beta}\, \bigg[\rm{1}+o(1)\bigg]\,,
\]
where $o(1)$ stands for a quantity that tends to zero as
$\rho\to\infty$ and $\Gamma$ is the Gamma function. Therefore, we infer that the
eigenfunction (\ref{Phic1c2})  will not be square-integrable at infinity, with respect to the corresponding radial measure (\ref{product1}), unless
$\al$ is a nonnegative integer $n$, so that the hypergeometric series appearing in the definition of 
$M^\beta_{-\alpha}(\rho)$ collapses to a finite sum, yielding the generalized Laguerre polynomials  $L^{\beta-1}_{\alpha}(\rho)$ of degree $\alpha=n$. Moreover, while $M^\beta_{-\alpha}(\rho)$ remains bounded at $0$, 
$U^{\beta}_{-\alpha}(\rho)$ diverges badly at the origin, so we must take $c_1=0$.

Hence, by substituting  $\alpha=n$    in~(\ref{defs}) 
we obtain the quadratic equation
\be
\la^2 E^2 + 2  \left(k \la+\hbar^2\left(n+l+\frac{N-1}{2}\right)^2\right) E  + k^2 =0,
\label{lanlan}
\ee
which we can readily solve to get
the following explicit expression for the discrete eigenvalues of $\hamG$: 
\be
E_{n,l} =\frac{-k^2}{\hbar^2\left(n+l+\frac{N-1}{2} \right)^2+k \la + \sqrt{\hbar^4\left(n+l+\frac{N-1}{2}\right)^4+2\,\hbar^2k \la \left(n+l+\frac{N-1}{2}\right)^2}} \, .
\label{lan}
\ee

We remark that the positive sign in the square root is chosen in such a manner that the limit $\eta\to 0$ exists and returns the non-deformed discrete spectrum
(\ref{energy}) and, moreover, 
 this result   ensures that the new coupling constant  $K$ (\ref{formK})  remains positive 
  for all   values of $n,l$. In this respect, notice also that  the equation (\ref{lanlan}) can  be rewritten in terms of $K$    in the form
\be
E=-\frac{K^2}{2 \hbar^2 \left(n+l+\frac{N-1}{2} \right)^2}\,,
\label{eqK}
\ee
to be compared with (\ref{energy}).

Then the eigenstates $\Phi(r)$ for $\hat{\mathcal{H}}$ can straightforwardly be obtained, in terms of $K$, by introducing  $\rho$ and $\beta$   given in (\ref{defs}),  $\alpha=n$ and the relation 
(\ref{eqK}) in the eigenfunction (\ref{Phic1c2}) with $c_1=0$. Next, the 
  similarity transformation~\eqref{lk} yields the eigenstates for $\hamG$; namely
\be
\Phi_{{\rm c}}(r) =\left(1+\frac{\la}{r} \right)^{\frac{2-N}{4}} r^l \exp\left( - \frac{K r}{ \hbar^2\left(n+l+\frac{N-1}{2}\right)}  \right)   \,L_n^{2l+N-2}\left( \frac{2 K r}{\hbar^2\left(n+l+\frac{N-1}{2}\right)}\right) ,
\label{egg}
\ee
where one must keep in mind that, in fact,  $K$ depends on $\eta$ and $E_{n,l}$ (\ref{lan}).

In view of the asymptotic behavior of the effective potential,
standard results for one-dimensional differential operators
show~\cite[Theorem XIII.7.66]{DS88} that the continuous spectrum for $\hamG$ is
the positive real line and that there are no eigenvalues embedded in
the continuum. Hence we can now   summarize all the above  results as follows.

\begin{theorem}\label{T.eigen} Let $\hamG$ be the quantum Hamiltonian  \eqref{kn} with $k>0$ and $\eta>0$. Then: 

\noindent
(i) The continuous spectrum of $\hamG$ is given by $[0,\infty)$. Moreover, there are no embedded eigenvalues and the singular spectrum is empty. 

\noindent
(ii)   $\hamG$ has an infinite number of eigenvalues $E_{n,l}$, depending only
on the sum $(n+l)$ and accumulating at $0$.

\noindent
(iii)  The eigenvalues of $\hamG $ are of the form~\eqref{lan} and $\Psi_\LB= \Phi_{{\rm c}}(r)Y(\boldsymbol{\te})$, determined by~\eqref{egg}, is eigenfunction of $\hamG $ with eigenvalue $E_{n,l}$. 

\end{theorem}

\begin{figure}
\begin{center}
\includegraphics[height=8.2cm]{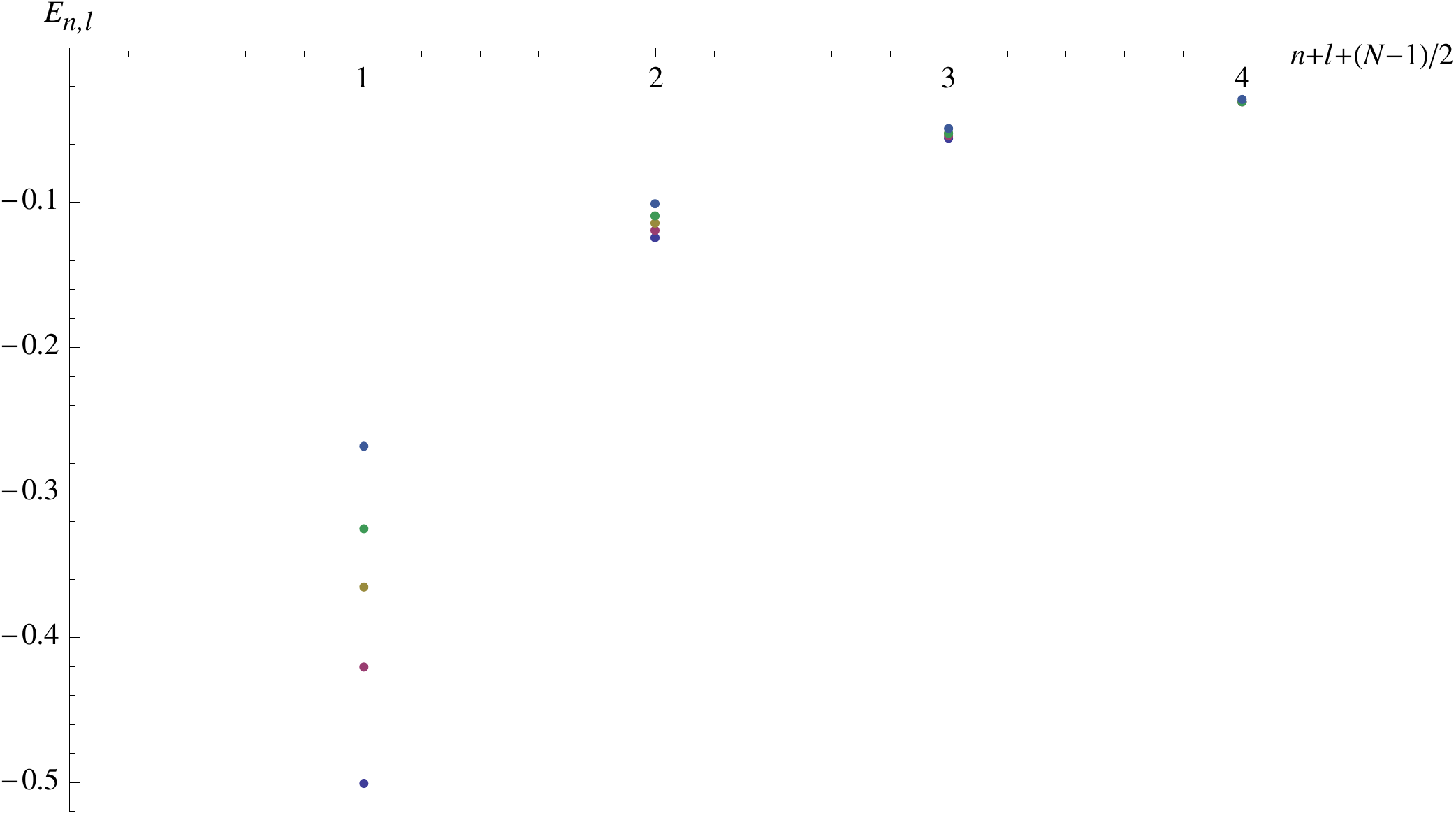}
\caption{Discrete spectrum  (\ref{lan}) for the fundamental and the three first excited states of the Hamiltonian $\hamG$ when $\eta=\{0, 0.2, 0.4, 0.6, 1\}$ with $\hbar=k=1$ and $N\geq 3$. Note that the effect of the $\eta$ deformation is quite strong for the fundamental state, since it comes from the shift $r\to r+\eta$ in the usual Coulomb potential. 
 \label{figure6}}
\end{center}
\end{figure}

Notice that, in particular, the bound states of this system satisfy
\be
  \lim_{n,l\to \infty}E_{n,l}= 0 ,\qquad  \lim_{\mathfrak{n}\to \infty}(E_{\mathfrak{n}+1}-E_\mathfrak{n})=0, \qquad \mathfrak{n} =n+ l.
  \nonumber
\ee
As it can be appreciated in
Figure~\ref{figure6}, deviations from the spectrum of the quantum
Coulomb problem are significant for the low energy states, since the
effect of the deformation is essentially a shift $r\to r+\la$ in the
standard Coulomb potential. In fact,  in the limit $\eta\to
0$ of~\eqref{lan} the well-known formula for the standard
Coulomb eigenvalues $ E^0_{n,l}$ (\ref{energy}) is recovered.
The  deformation on the spectrum can be better appreciated
through its power series expansion in $\eta$ given by
\begin{equation}
\nonumber
E_{n,l} =E^0_{n,l}  + \eta\,\frac{ k^3}{2 \hbar^4 \left(n+l+\frac{N-1}{2}\right)^4} 
-\eta^2\, \frac{5 k^4}{8 \hbar^6 \left(n+l+\frac{N-1}{2}\right)^6} + O(\eta^3) .
\end{equation}

 %%%%%%%%%%%%%%%%%%%%%%%%%%%%%%%%%%%%%%%%%%%%%%%%

%%%%%%%%%%%%%%%%%%%%%%%%%%%%%%%%%%%%%%%%%%%%%%%%

\subsect{A formal Lie-algebraic derivation of the spectrum}

Here we will show how the symmetry Lie algebra generated by the
constants of motion strongly suggests that the eigenvalues of the
system (for $k>0$ and $\eta>0$) are indeed of the form~\eqref{lan}  which we
rigorously have just found. In this
subsection we will present a formal algebraic derivation of the
spectrum, although the reader is advised to take it with a pinch of salt, since
subtle technical issues related to the domain of the operators make
this derivation non-rigorous; indeed, we shall see in the following
subsection that, for this reason, the case of negative $k$ and/or $\eta$
does not fit into this algebraic framework.

Let us start off with  the sum of the squared Runge--Lenz operators
introduced in (\ref{cc}), which turns out to be
\be
\hat{\mathbf{R}}^2= \sum_{i=1}^N \hat{\mathcal{R}}_i^2 = 2 \hat{\mathcal{H}}  \left(\hat{{\mathbf L}}^2 + \hbar^2 \frac{(N-1)^2}{4} \right) + \left(\eta \hat{\mathcal{H}}+k\right )^2 ,
\label{quantumpauli}
\ee
which is  the  quantum counterpart of the relation (\ref{funct}).
Now, if similarly to the classical case~\eqref{clasA}, we introduce the following operators (that can be only defined on the set of eigenfunctions of $\hat{H}_\eta$)
\begin{equation}
\hat{J}_{0i} = \frac{\hat{\cal R}_i}{\sqrt{- 2 \hat{\cal H}}}  ,
\nonumber
\end{equation}
then the equation (\ref{quantumpauli})  becomes formally
\begin{equation}
\sum_{i=1}^N \hat{J}_{0i}^2 + \hat{{\mathbf{L}}}^2  + \frac{\hbar^2 (N-1)^2}{4} = - \frac{(\eta \hat{\cal H} + k)^2 }{2 \hat{\cal H}},
\label{paul}
\end{equation}
where $\sum_{i=1}^N \hat{J}_{0i}^2 + \hat{{\mathbf{L}}}^2 $ is the
quadratic Casimir operator of the $\mathfrak{so}(N+1)$ Lie
algebra. Therefore, if this operator has its usual domain, its eigenvalues are
\be
\hbar^2 \mathfrak{n} (\mathfrak{n} + (N+1) -2) = \hbar^2 \mathfrak{n} (\mathfrak{n} + N - 1).
\nonumber
\ee
This implies that the eigenvalues of  the l.h.s.~of equation (\ref{paul})  
 should be $\hbar^2 (\mathfrak{n} + \frac{N-1}{2})^2$. Now, if we act
 with~\eqref{paul} onto an eigenfunction of $\hat{\cal H}$ with
 eigenvalue $E$, we formally get
\begin{equation}
- \frac{( \eta E+k )^2 }{2 E} = \hbar^2 \left(\mathfrak{n} + \frac{N-1}{2}\right)^2 ,
\nonumber
\end{equation} 
which is just the equation (\ref{lanlan}) (and also~\eqref{eqK}) for the energies where
$\mathfrak{n} = n + l$.

\subsect{The remaining cases}

For the sake of completeness, let us qualitatively describe the spectrum of the system for the remaining possible
values of $\eta$ and $k$.

\subsubsection*{The case $\eta>0$, $k<0$}

In this case the manifold is again
$\cM=\RR^N\backslash\{\mathbf{0}\}$, but the potential is
repulsive. The Hamiltonian $ \hat H_{\LB,l}$ can be again defined as the Friedrichs
extension of the action of the differential operator on $C^\infty_0(\RR^+)$. In view of our formulas for the quantum effective potential
on the space of degree--$l$ spherical harmonics, an easy application
of~\cite[Theorem XIII.7.66]{DS88} shows that the spectrum of $ \hat
H_{\LB,l}$ is purely absolutely continuous and given by $[0,\infty)$.

\subsubsection*{The case $\eta<0$, $k>0$}

Now the manifold is $\cM=\{\bq \in\RR^N: |\bq|>|\eta|\}$ and the
potential is attractive, so one expects to have eigenvalues. The Hamiltonian $ \hat H_{\LB,l}$ is defined
as the Friedrichs extension of the operator with domain
$C^\infty_0((|\eta|,\infty))$; its continuous spectrum is $[0,\infty)$
and again there is no embedded eigenvalues or singular continuous
spectrum.

Therefore, the eigenvalues we are looking for would be those values of $E$ for which the function $\Phi$~\eqref{Phic1c2} is square-integrable at infinity, with respect to
the induced radial measure, and such that this satisfies the boundary condition
$\Phi(|\eta|)=0$. In principle, one has  the values of $E<0$ for which the confluent hypergeometric 
  function $U$ satisfies the boundary condition
\[
U_{-\al}^{2l+N-1}\Big(\frac{2|\eta|\sqrt{-2E}}\hbar\Big)=0\,,
\]
where $\al$ also depends on $E$ via~\eqref{defs}.
These eigenfunctions correspond to set  $c_2=0$ in~\eqref{Phic1c2}.
This   kind of eigenvalues cannot be computed in closed form, but
it is not hard to see that one can take specific values of $k$ and
$\eta$ where there are indeed eigenvalues that satisfy these
conditions.

However, we remark that to obtain in a rigorous manner the eigenvalues for $\Phi$ with $c_2\ne 0$, that is,    
of the type $E_{n,l}$ (\ref{lan}),   is not straightforward at all and 
 a deeper analysis  of this delicate case is  deserved.

\subsubsection*{The case $\eta<0$, $k<0$}

Now the manifold is $\cM=\{\bq \in\RR^N: |\bq|>|\eta|\}$ and the
potential is repulsive. The Hamiltonian $ \hat H_{\LB,l}$ is defined
as the Friedrichs extension of the operator with domain
$C^\infty_0((|\eta|,\infty))$ and the spectrum, which consists of the
positive real line, is purely absolutely
continuous.

%%%%%%%%%%%%%%%%%%%%%%%%%%%%%%%%%%%%%%%%%%%%%%%%%%%

\sect{Concluding remarks}
 
In this paper we have presented a new exactly solvable quantum system in $N$ dimensions, that has been obtained as the maximally superintegrable quantization of the Hamiltonian~\eqref{aa}. Such quantization has been performed by making use of the conformal Laplacian prescription and its equivalence with the so-called ``direct Schr\"odinger" one. It is worth mentioning that, as it was shown in~\cite{annals}, both quantization approaches  can also be  related to the {\em position-dependent mass} quantization (see, for instance~\cite{Roos, mass2, Quesnea})
\begin{equation}
\nonumber
\hat{\mathcal{H}}_{\rm PDM} = -\frac{\hbar^2}2  \nabla\cdot \frac{1}{f(r)^2}\, \nabla +\mathcal U(\bq)\,   ,
\end{equation}
where $f(r)$ is the conformal factor of the metric (\ref{aacc}).

It is also important to remark that, although the potential of the system~\eqref{aa}, namely
$$
{\cal U}(\bq)=-\frac{\kk}{\la + |\bq|}
\label{pe}
$$
can be interpreted as an $\la$-deformation of the Coulomb problem on the $N$D Euclidean space,  ${\cal U}$ is by no means  superintegrable (and, therefore, exactly solvable) on such flat space. It turns out that in order to recover a superintegrable system containing this deformed Coulomb potential, the kinetic energy term has also to be  $\la$-deformed, and the outcome of this deformation is just a metric associated with the Taub--NUT space whose curvature~\eqref{ac} is again controlled by the deformation parameter $\la$. In this sense, the Hamiltonian~\eqref{aa} is a quite singular system in which the {\em same} parameter $\la$ plays both a dynamical role (in the potential term) and a geometric one (in the kinetic energy), and both roles have to be exactly tuned  in order to allow superintegrability to arise.

Therefore, the system~\eqref{aa} and its superintegrable conformal Laplacian quantization ${\hat{\cal H}}_{{\rm c} }$~\eqref{kn} can be indeed considered a deformation (in the two abovementioned dynamical and geometric senses) of the Euclidean Coulomb system, and such deformed quantum system can be fully solved by making use of the very same techniques used in the former. Nevertheless, we recall that the system~\eqref{aa} {cannot} be properly called the Coulomb problem on the Taub-NUT space with metric (\ref{aacc}). In fact, the geometrical definitions  of the  intrinsic  Coulomb ${\cal U}_{\rm C}$ and oscillator ${\cal U}_{\rm O}$ potentials  on   a  spherically symmetric space   are,  respectively, given by (see~\cite{annals2} for details)
$$
{\cal U}_{\rm C}(r) : =\int^r\frac{\dd r'}{r'^2f(r')} ,\qquad {\cal U}_{\rm O}(r) :=\frac  1{ {\cal U}_{\rm C}(r)^2} ,
$$
up to multiplicative and additive constants, 
where $f(r)$ is the conformal factor of the metric (\ref{aacc}).
Now, if we apply these expresions to the Taub-NUT space whose conformal factor is~\eqref{ad} we find that, whenever $\eta\ne 0$, the potential $\cal U$ in (\ref{aa}) defines an {\em intrinsic oscillator} potential ${\cal U}_{\rm O}$ on  the curved space $\cM$ with metric ({\ref{metr}) that would be given by  (see~\cite{JPCS} for more details) 
$$
{\cal U}_{\rm O}=C\frac{  r}{ \la + r}+D,
$$
where $C$ and $D$ are real constants. Hence, if we set  $C=k/\eta$ and $D=-C$, we get that ${\cal U}_{\rm O}\equiv {\cal U}$.
From a physical viewpoint,  this can be understood in the sense that the potential ${\cal U}$ takes a finite value on $r=0$ (see Figure 4), which means that the characteristic Coulomb potential singularity at the origin has been suppressed by the deformation.

%%%%%%%%%%%%%%%%%%%%%%%%%%%%%%%%%%%%%%%%%%%%%%%%%%%

\section*{Acknowledgments}

This work was partially supported by the Spanish MINECO through the Ram\'on y Cajal program (A.E.) and under grants  MTM2010-18556 (A.B and F.J.H.), AIC-D-2011-0711 (MINECO-INFN)  (A.B, F.J.H.~and O.R.) and   FIS2011-22566 (A.E.), by  the ICMAT Severo Ochoa under grant SEV-2011-0087 (A.E.),  by Banco Santander-UCM under grant GR35/10-A-910556 (A.E.), and by  a postdoctoral fellowship  by the Laboratory of
Mathematical Physics of the CRM, Universit\'e de Montr\'eal (D.R.).

 %%%%%%%%%%%%%%%%%%%%%%%%%%%%%%%%%%%%


\begin{thebibliography}{10}\frenchspacing

\small


\bibitem{IK94}
T.  Iwai, N.  Katayama,    {J.. Phys. A: Math. Gen.}  {27} (1994)   3179.





\bibitem{Ma82}
N.S. Manton, {Phys. Lett. B} {110} (1982) 54.

\bibitem{AH85}
M.F. Atiyah,  N.J.  Hitchin, {Phys. Lett. A} {107} (1985) 21.


\bibitem{GM86}
G.W. Gibbons, N.S. Manton, {Nucl. Phys. B} {274} (1986) 183.


\bibitem{FH87}
L.G. Feh\'er,  P.A.  Horv\'athy, {Phys. Lett. B} {183}  (1987) 182.


\bibitem{GR88}
G.W. Gibbons, P.J.  Ruback, {Comm. Math. Phys.} {115} (1988) 267.




\bibitem{IK95}
T. Iwai,  N. Katayama, {J. Math. Phys.} {36} (1995) 1790.


\bibitem{uwano}
T. Iwai,  Y.  Uwano, N.  Katayama,  J. Math. Phys. 37 (1996) 608.


 \bibitem{BCJ} D. Bini, C. Cherubini, R.T. Jantzen,    {Class. Quantum Grav.}  {19}  (2002)  5481.

\bibitem{BCJM} D. Bini, C. Cherubini, R.T. Jantzen, B. Mashhoon,   {Class. Quantum Grav.}  {20}    (2003) 457.

 \bibitem{GW07}
G.W. Gibbons, C.M. Warnick,  {J. Geom. Phys.} {57} (2007) 2286.


\bibitem{JL} J. Jezierski, M. Lukasik,    {Class. Quantum Grav.}  {24}  (2007)  1331.



\bibitem{sigma}
   A.   Ballesteros,  A. Enciso, F.J. Herranz, O. Ragnisco, D. Riglioni,   SIGMA 7 (2011) 048.


\bibitem{darbouxiii}    A. Ballesteros, A. Enciso, F.J. Herranz, O. Ragnisco, D. Riglioni,      {Phys. Lett. A}   {375} (2011) 1431.



\bibitem{annals}
   A. Ballesteros, A. Enciso, F.J. Herranz, O. Ragnisco, D. Riglioni,   {Ann. Phys.} {326} (2011) 2053. 





\bibitem{Ko72}
G. Koenigs, in:   {{L}e{\c{c}}ons sur la th\'eorie g\'en\'erale des surfaces}
  vol.  4,  ed.   G. Darboux, Chelsea, New York,  1972, p. 368.
  
\bibitem{KKMW03}
E.G.  Kalnins,  J.M. Kress, W. Jr. Miller,  P. Winternitz,
 {J. Math. Phys.} {44} (2003)  5811.



\bibitem{JPCS}
   A. Ballesteros, A. Enciso, F.J. Herranz, O. Ragnisco, D. Riglioni,   {J. Phys: Conf. Series} {474} (2013) 012008. 


  
   \bibitem{Baer}  C. Baer, M. Dahl, {Geom. Funct. Anal.} 13 (2003) 483. 
 

   
   \bibitem{MRS} J.P Michel, F. Radoux, J. Silhan, SIGMA 10 (2014) 016.

 





\bibitem{Wa84}
  R.M. Wald, {General Relativity}, The University of Chicago Press, Chicago, 1984.


\bibitem{Landsman}
   N.P. Landsman, {Mathematical Topics Between Classical and Quantum Mechanics}, Springer, New York (1998).


\bibitem{Liu}
   Z.J. Liu, M. Qian, {Trans. Amer. Math. Soc.}   {331} (1992) 321.



\bibitem{annals2}
   A. Ballesteros, A. Enciso, F.J. Herranz, O. Ragnisco,    {Ann. Phys.} {324} (2009) 1219. 



\bibitem{REFS1} R.S. Strichartz, J. Funct. Anal. 52 (1983) 48. 


 


  \bibitem{DS88}
  N. Dunford, J.T. Schwartz, {Linear Operators II}, Wiley, New York, 1988.


   \bibitem{Roos}
O. von Roos,     {Phys. Rev. B}  {27}  (1983) 7547.
   
\bibitem{mass2}
J.M. Lévy-Leblond, Phys. Rev. A 52 (1995) 1845. 

\bibitem{Quesnea}
C. Quesne, V.M. Tkachuk, J. Phys. A: Math. Gen. 37 (2004) 4267.



 
 
    
  

 



\end{thebibliography}
\end{document}